\documentclass[aps,prl,twocolumn,superscriptaddress,showpacs]{revtex4}
\usepackage{graphicx}

\begin{document}
\newcommand{\ud}{{\mathrm d}}
\newcommand{\sech}{\mathrm{sech}}

\title{High-frequency effects in the FitzHugh-Nagumo neuron model}


\author{David Cubero}
\thanks{Corresponding author. Email address: dcubero@us.es}
\affiliation{F\'{\i}sica Te\'orica, Universidad de Sevilla, Apartado de Correos 1065, Sevilla 41080, Spain}
\author{J. P. Baltan\'as}
\email[]{baltanas@us.es}
\affiliation{Departamento de Matem\'aticas y F\'{\i}sica Aplicadas y Ciencias de la Naturaleza, Universidad Rey Juan Carlos, Tulip\'an s/n, M\'ostoles 28933, Madrid, Spain}
\author{Jes\'us Casado-Pascual}
\email[]{jcasado@us.es}
\affiliation{F\'{\i}sica Te\'orica, Universidad de Sevilla, Apartado de Correos 1065, Sevilla 41080, Spain}


\date{\today}

\begin{abstract}
The effect of a high-frequency signal on the FitzHugh-Nagumo excitable model is analyzed. We show that the firing rate is diminished as the ratio of the high-frequency amplitude to its frequency is increased. Moreover, it is demonstrated that the excitable character of the system, and consequently the firing activity, is suppressed for ratios above a given threshold value. In addition, we show that the vibrational resonance phenomenon turns up for sufficiently large noise strength values.   
\end{abstract}


\maketitle
Non-linear noisy systems have been studied with ever growing interest due to its applicability to the modelling of a great variety of phenomena of relevance in physics, chemistry and the life sciences \cite{gamjun98,han02}. 
Perhaps, the simplest non-linear noisy system studied in the literature is the bistable system, which has been used successfully to illustrate the phenomenon of stochastic resonance. 
In addition, the spiking activity of a neuron has been theoretically studied using nonlinear excitable noisy models \cite{koch99}, among them, the FitzHugh-Nagumo (FHN) system \cite{fit61} being one of the most utilized due to its simplicity. Besides, networks of FHN units have been considered as simplified models useful in the description of both the cardiac tissue \cite{alosag03,grajal96,panfilov97,win91} and reaction-diffusion chemical systems \cite{kapral95,win91}.  

Recently, it has been shown, both theoretically and experimentally, that the addition of a high-frequency (HF) signal results in an improvement of the stochastic resonance in a bistable optical system \cite{chigia05}. Nevertheless, many experimental studies have suggested that HF (non-ionizing) fields may damage several structural and functional properties of neuronal membranes in single cells, as well as to cause a number of negative physiological effects on typically excitable media such as the cardiac tissue \cite{ahlber98}. Besides, it has been found that certain HF signals are able to suppress the steady directed motion in a ratchet model \cite{bormar05}. Thus, HF signals seem to play a two-fold role, positive or negative, depending on the non-linear system under consideration.

In this paper we study the influence of a HF field on a rather simple excitable model, such as the archetypal FHN system.
This system is governed by the following equations \cite{footnote}
\begin{eqnarray}
\varepsilon \dot{v}&=&v -v^3-w+\xi(t)+S(t)+\Gamma(t)
\label{eq:fnv} \\
\dot{w}&=&\gamma v-w+b. \label{eq:fnw}
\end{eqnarray}
In the context of neurophysiology, $v(t)$ is called the voltage
variable, and $w(t)$ the recovery variable. Here $S(t)$ is an external forcing of period $T$, $\xi(t)$ a noisy term, and $\Gamma(t)$ is an added HF signal which will be specified later on. The ``neuronal'' noise  $\xi(t)$ is assumed to be an unbiased Gaussian white noise with autocorrelation function $\langle \xi(t)\xi(s)\rangle=2D\delta(t-s)$. The values of the model parameters define the dynamical regimes of the system, and are discussed below.    

Let us focus our attention in the deterministic FHN model ($D=0$) in the absence of HF signal ($\Gamma(t)=0$). In the case of a time-independent external
signal $S(t)=S_0$, the excitable character of the FHN model relies on the
existence of a threshold value of this signal, $S_\mathrm{H}$, at which a Hopf
bifurcation takes place. A simple linear stability analysis
\cite{fit61} shows that the condition $\varepsilon<1$ is necessary to assure the
existence of such a threshold $S_\mathrm{H}$.  In fact, it is usually assumed that $\varepsilon\ll 1$ so that the voltage variable is much faster than the recovery variable. In addition, we must choose
$\gamma>1$ in order to guarantee that for subthreshold signals there exits a unique stable attractor \cite{fit61}. This is usually the case of interest in most applications of the model to real problems, including neuron dynamics. In this situation, for $S_0<S_\mathrm{H}$ the system is quiescent, i.e., the
attractor is a globally stable fixed point $(v_0,w_0)$. When the threshold value $S_\mathrm{H}$ is exceeded, the fixed point becomes unstable and a stable limit cycle appears. As a consequence, when $v(t)$ is viewed as a function of time, a sequence of ``spikes'' is observed, each spike resembling an action potential as detected in real neurons.

\begin{figure}
\includegraphics[width=7cm,angle=-90]{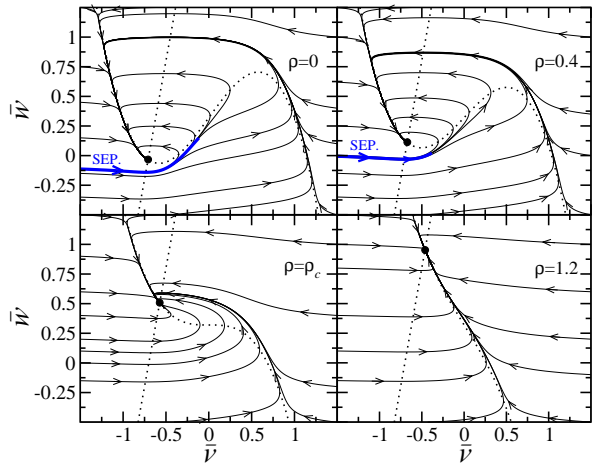}
\caption{
\label{fig:fnphase} 
(Color online) Deterministic time evolution in phase space $(\bar{v},\bar{w})$ for different representative initial conditions and a stationary subthreshold external force $S(t)=S_0=0.32$. From top-left to bottom-right: $\rho=0$ (no HF signal); $\rho=0.4$; $\rho=\rho_c=0.80829\ldots$; and $\rho=1.2$. The stable attractors are depicted by filled solid circles. The nullclines [i.e., $\gamma \bar{v}-\bar{w}+b=0$ and $c(\rho)\bar{v}-\bar{v}^3-\bar{w}+A=0$] are represented by dotted lines. Top panels show the separatrix between nonexcitable and excitable behaviors by  thick (blue) solid lines. (All quantities in dimensionless units.)
 }
\end{figure}

\begin{figure}
\includegraphics[width=7cm,angle=-90]{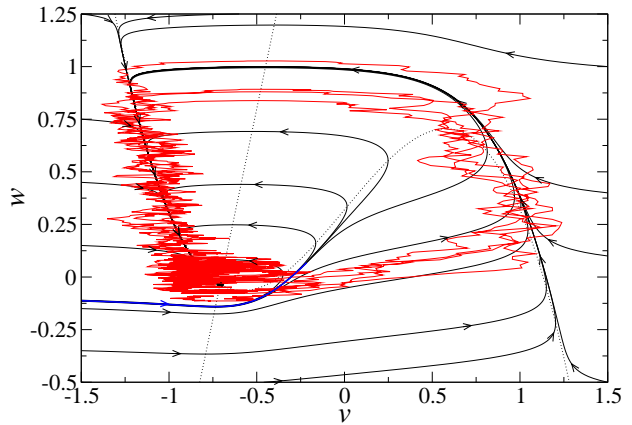}
\caption{
\label{fig:r0_vw} 
(Color online) Typical phase space trajectory for the same parameter values as in the top-left panel of Fig.~\ref{fig:fnphase} and a noise strength $D=5\times 10^{-4}$. (All quantities in dimensionless units.)}
\end{figure}

\begin{figure}
\includegraphics[width=7cm,angle=-90]{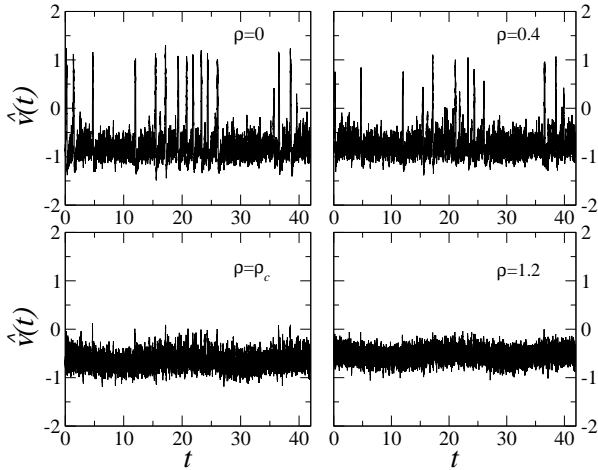}
\caption{
\label{fig:rc} 
Typical trajectories $\hat{v}(t)$ for a noise strength $D=5\times 10^{-4}$ and an external force $S(t)=A\cos(\Omega t)$,  with $A=0.32$ and $\Omega=0.3$. The HF signal is $\Gamma(t)=N\Omega \rho\varepsilon  \cos(N \Omega t)$, with $N=4000$ and the same values of $\rho$ as in Fig.~\ref{fig:fnphase}. (All quantities in dimensionless units.)}
\end{figure}

In the top-left panel of Fig.~\ref{fig:fnphase}, we show several representative phase-space trajectories obtained by solving the deterministic FHN equations for initial conditions at the edges of the frame. The legend $\rho=0$ corresponds to $\Gamma(t)=0$, as will be clarified below. We have used the standard parameter values \cite{lon93},  $\varepsilon=0.02$, $\gamma=4$, and $b=2.8$, which corresponds to an excitable regime. The external forcing is stationary with $S_0=0.32$. Since this is a subthreshold value [the threshold value is given by Eq.~(\ref{eq:sh}) with $\rho=0$], all trajectories end up at the stable attractor, which is depicted by a solid filled circle. With a thick (blue) solid line we have plotted the separatrix which separates a region with excitable trajectories (those with a large excursion in phase space) from a region with non-excitable ones (shorter excursion). The excitable paths are reminiscences of the above mentioned stable limit cycle, which is only present in the superthreshold regime. In the presence of noise, the subthreshold dynamics consist of a random trajectory around the stable attractor, which eventually crosses the separatrix into the excitable region (see Fig.~\ref{fig:r0_vw}). These excitations appear as random spikes in the variable $v(t)$, as shown in the top-left panel of Fig.~\ref{fig:rc}. Notice that in this panel the external forcing $S(t)=A\cos(\Omega t)$ is time-dependent. When the external forcing varies slowly compared with the characteristic time-scales of the FHN model, the above qualitative scheme remains valid if one considers a sequence of diagrams corresponding to the instantaneous values of $S(t)$.  

Let us now study the effect of a HF signal. In this paper, we assume that ${\Gamma}(t)$ takes the form:
\begin{equation}
{\Gamma}(t)=N\Omega {r} \cos(N \Omega t+\varphi),
\label{eq:cs}
\end{equation}
where $\Omega=2\pi/T$, $\varphi$ is an arbitrary phase, $N$ is a large
integer and ${r}$ is the ratio of the amplitude of the HF signal to its frequency. We are interested in situations in which the parameters $N\Omega {r}$ and $N \Omega$ are much larger than the rest of the parameters in the problem. In this sense, we will say that ${\Gamma}(t)$ is a strong, high frequency monochromatic force.
This situation corresponds formally to taking the limit
$N\rightarrow \infty$ while keeping the ratio $r$ fixed. The assumption of a strong signal is necessary if the high frequency signal is to have any effect.

In this limit it is clear from Eq.~(\ref{eq:fnv}) that the stochastic process ${v}(t)$ is "fast", since its time
derivative is of order $N$.  A "slow" stochastic process, $\hat{v}(t)$, can be defined after extracting the "fast"
dependence from ${v}(t)$,
\begin{equation}
\hat{v}(t)={v}(t)-{\rho}\sin(N\Omega t+\varphi),
\label{eq:vhatdef}
\end{equation}
where $\rho=r/\varepsilon$.
Using Eq.~(\ref{eq:vhatdef}) in (\ref{eq:fnv})--(\ref{eq:fnw}), it is easy to check that the time derivatives of both $\hat{v}(t)$ and $\hat{w}(t)=w(t)$ are of order 1 as $N\rightarrow \infty$. Therefore, a large number of oscillations of the function $\rho\sin(N\Omega t+\varphi)$ takes place before a significant change of  $\hat{v}(t)$ and $\hat{w}(t)$ occurs. As a consequence, in this limit, their dynamics become independent of the particular value of the phase $\varphi$, and the resulting equations of motion can be simplified by averaging over all possible values of this phase. Then, the following equations are obtained
\begin{eqnarray}
\varepsilon \dot{\bar{v}}&=&c(\rho)\bar{v}-\bar{v}^3+S(t)+\xi(t) \label{eq:vhat} \\
\dot{\bar{w}}&=&\gamma\bar{v}-\bar{w}+b. \label{eq:what}
\end{eqnarray}
where $c(\rho)=1-3\rho^2/2$, $\bar{v}(t)=(2\pi)^{-1}\int_0^{2\pi}d\varphi\,\hat{v}(t)$, and $\bar{w}(t)$ is defined analogously.

A linear stability analysis of the deterministic version of Eqs.~(\ref{eq:vhat})--(\ref{eq:what}) for a time-independent external signal $S_0$, shows that the above mentioned condition $\varepsilon<1$, which assured the existence of the limit cycle, is replaced in this case by $\varepsilon<c(\rho)$. Thus, for all values $\rho \ge \rho_c=\sqrt{2(1-\varepsilon)/3}$ the fixed point is stable regardless of the value of $S_0$. This fact leads to the disappearance of the separatrix for subthreshold signals, as shown in the bottom panels of Fig~\ref{fig:fnphase}. As a consequence, the system looses its excitability and no spikes are fired. 

As mentioned before, this description is appropriate even in the presence of a slow time-dependent external signal. This case have been considered in Fig.~\ref{fig:rc}, where we present four typical trajectories $\hat{v}(t)$ for the same values of $\rho$ as in Fig.~\ref{fig:fnphase} and a slow time-dependent external forcing $S(t)=A\cos(\Omega t)$, with $A=0.32$ and $\Omega=0.3$. The remaining parameter values are $D=5\times 10^{-4}$ and $N=4000$. The bottom panels show vividly the vanishment of the spikes due to the suppression of the system excitability by the HF signal.

In addition, the above stability analysis also shows that for ratio values $\rho\le \rho_c$, the Hopf bifurcation threshold is given by
\begin{equation}
S_H=b-\left(\frac{c(\rho)-\varepsilon}{3}\right)^{1/2}[\gamma-c(\rho)]-\left(\frac{c(\rho)-\varepsilon}{3}\right)^{3/2}.
\label{eq:sh}
\end{equation}
Thus, the larger the ratio $\rho$, the larger the threshold $S_H$. Consequently, the limit cycle becomes less accessible in the presence of the HF signal. This suggests that for subthreshold signals the spike rate decreases as $\rho$ increases. This fact is corroborated by the numerical simulations.  

In order to quantify the effect of the HF signal in the FHN model, let us consider the dependence of the spectral amplification $\eta$ on the ratio $\rho$. Since we are interested in the firing dynamics, first we reduce the stochastic process $\hat{v}(t)$ to a spike train $\Theta(t)$ by using the procedure of Ref.~\cite{wiepie94}. That is, a rectangular pulse of height unity and time length 0.15 is generated whenever $\hat{v}(t)$ crosses a threshold value, chosen as $v_\mathrm{th}=0.5$, having previously crossed a reset value $v_\mathrm{r}=-0.5$. We have chosen the output variable $\hat{v}(t)$ given by  Eq.~(\ref{eq:vhatdef}), instead of $v(t)$, to avoid the counting of fake firings due to the rapid oscillations. Then, the spectral amplification associated to the process $\Theta(t)$ is defined as the ratio of the integrated power stored in the first delta-like spike of the output power spectral density to that stored in the corresponding delta-like spike of the input power spectral density. It can be shown \cite{junhan89} that this quantity is given by $\eta=4|M_1|^2/A^2$, where $M_1=\int_0^T dt \,\langle \Theta(t)\rangle e^{-i\,\Omega t}/T$.

\begin{figure}
\includegraphics[width=7cm,angle=-90]{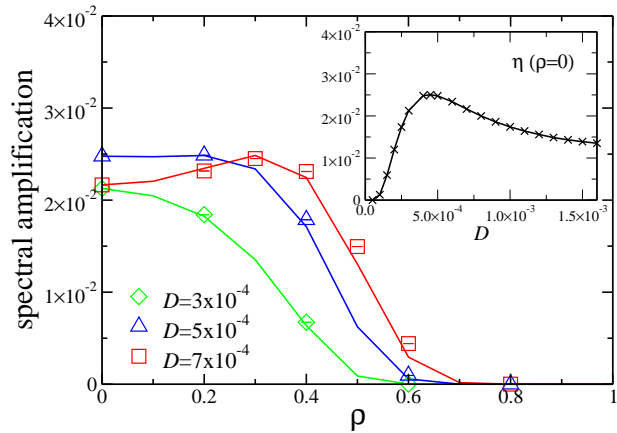}
\caption{
\label{fig:VR}
(Color online) Spectral amplification $\eta$ as a function of $\rho$. The noise strength values are:  
$D=3\times 10^{-4}$ (diamonds), $D=5\times 10^{-4}$ (triangles), and $D=7\times 10^{-4}$ (squares).
The values of $N$, $A$, and $\Omega$ are the same as in Fig.~\ref{fig:rc}. Solid lines depict the results obtained from $\bar{v}$ ($N\rightarrow\infty$). Inset shows $\eta$ vs.~$D$ in the absence of HF signal. (All quantities in dimensionless units.)}
\end{figure}

In Fig.~\ref{fig:VR}, we present the dependence
of $\eta$ on $\rho$ for three values of the noise strength. Solid lines
depict the simulation results using
Eq.~(\ref{eq:vhat})--(\ref{eq:what}) (the case $N\rightarrow\infty$),
whereas points represent the data with $N=4000$. Notice that $\eta$
vanishes for a given value of $\rho$ as a consequence of the
disappearance of the model excitability, the exact value of $\rho$
depending on the filter parameters $v_\mathrm{th}$ and
$v_\mathrm{r}$. For the lowest and the intermediate values of the
noise, $D=3\times 10^{-4}$ and $D=5\times 10^{-4}$, $\eta$ decreases
monotonously with $\rho$, although in the second case a plateau can be
observed for small values of this parameter. In contrast, for the
largest value of the noise strength, $D=7\times 10^{-4}$, $\eta$ peaks
at an optimal value of $\rho$. This non-monotonic behavior has also
been detected recently in bistable systems and has been coined
vibrational resonance (VR) \cite{lanmcc00,casbal04}. However, the
situation here is just the opposite to that observed in bistable
systems, because in the deterministic FHN model there is no VR effect,
and it only appears when the strength of the added noise is large
enough. 

VR in excitable systems has also been reported in
Ref.~\cite{ullzai03}. In contrast to our work, in that reference,
there is no clear-cut separation between the time-scales associated
with the HF signal, $(N\Omega)^{-1}$, and with the firing events. As a
consequence, they observe a different behavior. Besides, in
Ref.~\cite{ullzai03}, the HF signal is added to Eq.~(\ref{eq:fnw})
instead of to (\ref{eq:fnv}). Since $w(t)$ appears linearly in both
Eqs. (\ref{eq:fnv}) and (\ref{eq:fnw}), using the above techniques, it
is easy to prove that in the formal limit $N\rightarrow\infty$, the HF
signal would produce no effect.
  
In order to understand the behavior observed in Fig.~\ref{fig:VR}, let us consider first the
dependence of $\eta$ vs.~$D$ in the absence of HF signal ($\rho=0$). As shown in the inset of Fig.~\ref{fig:VR}, the spectral amplification displays a non-monotonic behavior,
with a maximum at an optimal noise strength $D_{\mathrm{opt}}$ at
which the coherence of the output signal is maximized. This is the
celebrated phenomenon of stochastic resonance, reported for the FHN model in
Refs.~\cite{lon93,wiepie94}. For $D<D_{\mathrm{opt}}$ the coherence of the output signal is less than
the optimal because the number of spikes is insufficient to optimize the coherence of the output
signal. In contrast, for $D>D_{\mathrm{opt}}$, the number of spikes is excessive, and an
appreciable number of them are fired out of phase with respect to $S(t)$, which degrades the
coherence. Diamonds in Fig.~\ref{fig:VR} correspond to a noise strength less than the optimal. Since the presence of the HF signal decreases further the number of spikes, $\eta$ must decrease monotonously as $\rho$ increases. On the other hand, when $D>D_{\mathrm{opt}}$ (squares in Fig.~\ref{fig:VR}), the decrease in the firing rate produced by the HF signal removes the incoherent extra spikes.  This fact brings the system
near the optimal operation regime, which results in a higher
value of $\eta$ than the corresponding to $\rho=0$. Of course, when
$\rho$ is increased further, the number of spikes decrease until total disappearance at a critical value of $\rho$, as discussed above.  Consequently, a non-monotonic behavior
of $\eta$ as a function of $\rho$ is expected for a fixed value of
$D>D_{\mathrm{opt}}$. Triangles in Fig.~\ref{fig:VR} correspond to an intermediate value of $D$, close to the optimal. 

In conclusion, we have shown that while a HF signal is able to improve the response of the FHN neuron system when the noise strength is larger than $D_{\mathrm{opt}}$, when $\rho$ is large enough, the HF signal inhibits completely the firing activity. This suggests a negative effect in the neuron activity for sufficiently strong HF signals which could be related to the functional biological damages caused by electro-pollution \cite{ahlber98}.

\begin{acknowledgments}

This research was supported by the Juan de la Cierva program of the Ministerio de Ciencia y Tecnolog\'{\i}a (D.C.), the Direcci\'on General de Ense\~nanza Superior of Spain by Projects Nos. BFM2002-03822 (J.C.-P.), the Junta de Andalucia (J.C.-P. and D.C.), and the Universidad Rey Juan Carlos under Project PPR-2004-38 (J.-P.B.). 
\end{acknowledgments}


\end{document}